# The influence of residual oxidizing impurities on the synthesis of graphene by atmospheric pressure chemical vapor deposition

Nicolas Reckinger[a,*], Alexandre Felten[a], Cristiane N. Santos[b], Benoît Hackens[b], and Jean-François Colomer[a]

[a]Research Center in Physics of Matter and Radiation (PMR), University of Namur, Rue de Bruxelles 61, B-5000 Namur, Belgium

[b]Université catholique de Louvain (UCL), Institute of Condensed Matter and Nanosciences (IMCN), Nanophysics Division (NAPS), Chemin du Cyclotron 2, 1348 Louvain-la-Neuve, Belgium

The growth of graphene on copper by atmospheric pressure chemical vapor deposition in a system free of pumping equipment is investigated. The emphasis is put on the necessity of hydrogen presence during graphene synthesis and cooling. In the absence of hydrogen during the growth step or cooling at slow rate, weak carbon coverage, consisting mostly of oxidized and amorphous carbon, is obtained on the copper catalyst. The oxidation originates from the inevitable occurrence of residual oxidizing impurities in the reactor's atmosphere. Graphene with appreciable coverage can be grown within the vacuum-free furnace only upon admitting hydrogen during the growth step. After formation, it is preserved from the destructive effect of residual oxidizing contaminants once exposure at high temperature is minimized by fast cooling or hydrogen flow. Under these conditions, micrometer-sized hexagon-shaped graphene domains of high structural quality are achieved.

[*] Corresponding author: Fax: +32 81 72 44 64
E-mail address: nicolas.reckinger@unamur.be (N. Reckinger).

# 1. Introduction

To be of use in industrial applications, large-area uniform graphene films of high structural quality have to be produced at low cost. In that context, catalytic chemical vapor deposition (CVD) shows great promise to fulfill this objective. The very first investigations of graphene formation on metals from gaseous hydrocarbons, dating back to 4-5 years ago, involved ruthenium [1], nickel [2,3], iridium [4] or copper [5]. In the meantime, several other metals have been added to that list: cobalt [6], platinum [7], molybdenum [8], etc. Still, presently, the most popular catalysts to grow graphene are copper and nickel because of their modest cost and availability. In their pioneering work, Li *et al.* [5] have evidenced that copper holds a great advantage over nickel in that it allows a self-limited growth resulting in monolayer graphene, while nickel often results in inhomogeneous films due to its much higher carbon solubility, making copper the favored catalyst. Since then, a plethora of publications have followed, investigating graphene synthesis in a wide range of conditions [9,10,11,12,13,14,15,16,17,18,19,20,21,22,23,24,25]. Various kinds of substrates have been used: commercial [5,9,10,12,13,14,16,18] or high purity copper foils [12,13], copper single crystals [15], thin copper films deposited onto thin layers grown or deposited on silicon wafers (silicon dioxide [17,18,19] or others [19]), copper thin films heteroepitaxially grown on sapphire [19,20,21,22] or MgO [23], melted copper [24,25]. Several pressure conditions have been considered: atmospheric [9,10,11,12,13,16,21,22] or low pressure [5,11,12,14,23], ultrahigh vacuum [11,15]. In terms of inexpensive and simple production of graphene, atmospheric pressure CVD (APCVD) is very likely the most attractive technique since it avoids pumping systems. However, the so-called "vacuum-free" APCVD reactors are kept most of the time under air. On the other hand, trace amounts of contaminants are always present in insufficiently pure gases and are the cause of graphene etching [26]. This raises the question of the purification of the atmosphere inside the tube and, more especially, the plausible presence of residual oxidizing species. Oxidative etching of graphene has indeed been reported previously and it is of great importance to avoid it since it considerably modifies both the structural and electrical properties of graphene through the introduction of defects and doping [26,27,28]. In the increasingly abundant literature about graphene growth by APCVD, that concern is very rarely tackled. Most of researchers making use of APCVD only mention that they flow Ar, $H_2$ or Ar/$H_2$ for some time before heating, without further details, except for Wu *et al.* [10] who report preparing their tube by a preliminary cycle of pumping and purges with Ar, implying a more expensive and complex equipment.

In this study, we report the growth of graphene by APCVD with methane on copper foils in a vacuum-free system. More specifically, the presence and the effect of residual oxidizing contaminants are indirectly investigated through the suppression of $H_2$ during the growth and/or cooling steps. For identical synthesis parameters, it is found that graphene is always obtained for a fast cooling (regardless of the gas mixture) while, in the case of a slow cooling, it is observed only with the addition of $H_2$ to Ar. This difference is explained by the inevitable presence of residual impurities, which etch the graphene film for long exposure times at high temperature.

## 2. Experimental details

### 2.1. Graphene growth

Before graphene growth, the copper foil (99.9% purity; 50-µm-thick) pieces are sonicated in acetone for 15 min, then in isopropanol for 15 min, and finally gently blown dry with nitrogen. Afterwards, superficial copper oxide is removed by a treatment with acetic acid (99.5% purity) at 35 °C for 10 min and the copper piece is blown dry with nitrogen without prior rinsing [29]. Immediately, the sample is deposited over a quartz boat introduced into a horizontal quartz reactor at room temperature. The reactor can be inserted/extracted into/from the furnace's hot zone rapidly. The synthesis process is summarized by the temperature-time diagram shown in Fig. S1. After sealing the reactor, ultrapure Ar (99.9999% purity) is flowed for 15 min with a rate of 2000 sccm under atmospheric pressure. Next, the quartz tube is inserted into the furnace (kept at 700 °C), 100 sccm of $H_2$ (99.9% purity) are added, the temperature is raised to 1000 °C, and the copper piece is annealed in these conditions for 1 h. Graphene is then grown by admitting 0.5 sccm of $CH_4$ (99.5% purity) for 15 min. All the previous parameters are identical for all the samples. The cooling is performed under 500 sccm of Ar with or without $H_2$, either rapidly (the tube is extracted manually outside the furnace) or slowly (5 °C/min between 1000 °C and 700 °C, then rapid cooling between 700 °C and room temperature).

## 2.2. Graphene transfer

Graphene is transferred onto $SiO_2$ (300-nm-thick)/Si wafers by the usual method based on polymethyl methacrylate (PMMA) [3]. First, a PMMA layer is spin-coated at 3000 rpm for 1 min over the front side of the graphene/copper sample and baked on a hot plate at 100 °C for 5 min. After protecting the PMMA, graphene grown on the back face is removed by oxygen plasma (50 W for 5 min). Then, the copper foil is immersed overnight in aqueous ammonium persulfate. The floating PMMA/graphene film is rinsed in distilled water, transferred to a $SiO_2$/Si piece and left to dry in air. A second PMMA layer is spin-coated over the first one in the same conditions. Finally, PMMA is removed by dipping the sample into acetone.

## 2.3. Characterization

Routine examination of the samples is made using an optical microscope (Olympus BX61), in the reflection mode, with a 60× objective. X-ray photoelectron spectroscopy (XPS) measurements are performed with an Escalab 250 Xi from Thermo. The spot size and pass energy are set to 200 µm and 20 eV, respectively. A monochromatized Al Kα X-ray source is used as photon source and photoelectrons are collected at an angle of 0° relative to the sample surface normal. Morphology and microstructure of graphene domains are examined with a Field Emission Scanning Electron Microscope (FE-SEM, JEOL JSM-7500F). The observation conditions are adapted from reference [30] by using mixed secondary and backscattered electron signals to clearly observe graphene (15% of backscattered electrons). The observation conditions are: working distance of 3 mm, accelerating voltage of 1 kV, emission current of 5 µA, low gentle beam mode with 0.2 kV applied to the specimen). Raman spectroscopy is performed at room temperature with a LabRam Horiba spectrometer with a laser wavelength of 514 nm. The laser beam is focused on the sample with a 100× objective (NA = 0.95) and the incident power is kept below 1 mW. Low and high resolution gratings are used in our measurements (150 and 1800 g/mm, respectively).

## 3. Results and discussion

To begin with, we set out to determine a set of appropriate parameters for graphene growth. The amount of $H_2$ during growth, the amount of $H_2$ during cooling, and the cooling rate are varied. The studied conditions are summarized in Table 1. For a fast and simple assessment of the growth conditions, we have made an extensive use of optical microscopy (OM). In order to facilitate the visualization of graphene on copper, the copper foils are baked on a heating plate at 150 °C for 5 min in air [31]. It results that bare copper is oxidized and graphene-covered copper is not, providing a very good contrast by means of OM.

| Sample list | | | |
|---|---|---|---|
| Sample name | $H_2$ for growth [sccm] | $H_2$ for cooling [sccm] | Cooling type |
| 100/10/S | 100 | 10 | slow |
| 100/0/S | 100 | 0 | slow |
| 100/10/F | 100 | 10 | fast |
| 100/0/F | 100 | 0 | fast |
| 0/10/S | 0 | 10 | slow |
| 0/0/S | 0 | 0 | slow |

**Table 1:** List of the samples, the varied parameters, and the final result. The abbreviations used for the sample names are self-explanatory.

In Fig. 1, OM images with the corresponding photographs of the entire samples (insets) illustrate the surface of two samples grown with $H_2$ and cooled slowly with Ar or Ar+$H_2$, before and after treatment on the heating plate. To facilitate the reading of the manuscript, the samples are dubbed 100/0/S and 100/10/S, respectively (see Table 1). For the sake of direct comparison, the OM images are taken exactly at the same spot, close to the center of the copper foil pieces. By naked eye only, it is difficult to detect any visible difference between the two samples before thermal treatment. With photography, however, it is apparent that sample 100/0/S is slightly redder (inset to Fig. 1a) compared with sample 100/10/S (inset to Fig. 1c). In addition, after baking in air, sample 100/10/S tarnishes a little (inset to Fig. 1d) while the other sample grows even redder (inset to Fig. 1b), sign that it has been oxidized. Since graphene offers protection to metal surfaces against oxidation [32], this means that the surface of sample 100/0/S must be graphene-free. This assertion is indeed supported by the

OM images displayed in Fig. 1a and b. Before hot-plate baking, sample 100/0/S is flecked with bluish-reddish stains all over its surface. By contrast, these speckles are not observed on sample 100/10/S (Fig. 1c). After annealing in air, the exposed copper surface of both samples is oxidized, turning to orange (Fig. 1b and d), while graphene-coated copper keeps its original hue. Strikingly, the two samples appear radically different: sample 100/0/S is almost entirely oxidized and scattered with small graphene dots while the copper surface of its *alter ego* is nearly completely covered by micrometer-sized graphene platelets. The partial coverage of sample 100/10/S explains its light tarnishing. Series of fringes due to ripples [33] in graphene can be plainly made out for many flakes (Fig. 1d). Upon careful *a posteriori* examination, the occurrence of such ripples can in fact already be observed before thermal treatment (see Fig. 1c). The formation of such ripples is related to dynamic instabilities at the interface of a carbon-catalyst binary system [33].

An identical comparison is also performed for two samples cooled rapidly, with or without $H_2$ (100/10/F and 100/0/F). Their visual aspect strongly resembles sample 100/10/S, with a copper surface scattered with graphene flakes (see Fig. S2a). Finally, two samples are grown without $H_2$. Both samples are cooled slowly, one with $H_2$ (0/10/S) and the other without (0/0/S). Each of them looks very similar to sample 100/0/S (see Fig. S2b and c). In the case of sample 0/10/S, it can be deduced that graphene growth is dramatically restrained without $H_2$, since the presence of $H_2$ during slow cooling should protect graphene, if grown, according to our previous observations. This is in contradiction with earlier reports in which defect-free graphene is successfully synthesized in the absence of $H_2$ during the growth [34,35]. At this point, two intermediary observations can be made from the previous findings. First, the slow cooling under Ar only (no $H_2$) is deleterious to graphene, as though it was somehow removed when exposed for too long to high temperatures in the absence of $H_2$, while graphene is preserved upon fast cooling whether $H_2$ is flowed or not. Second, the presence of $H_2$ during the growth seems to be a necessary requirement to obtain appreciable graphene coverage.

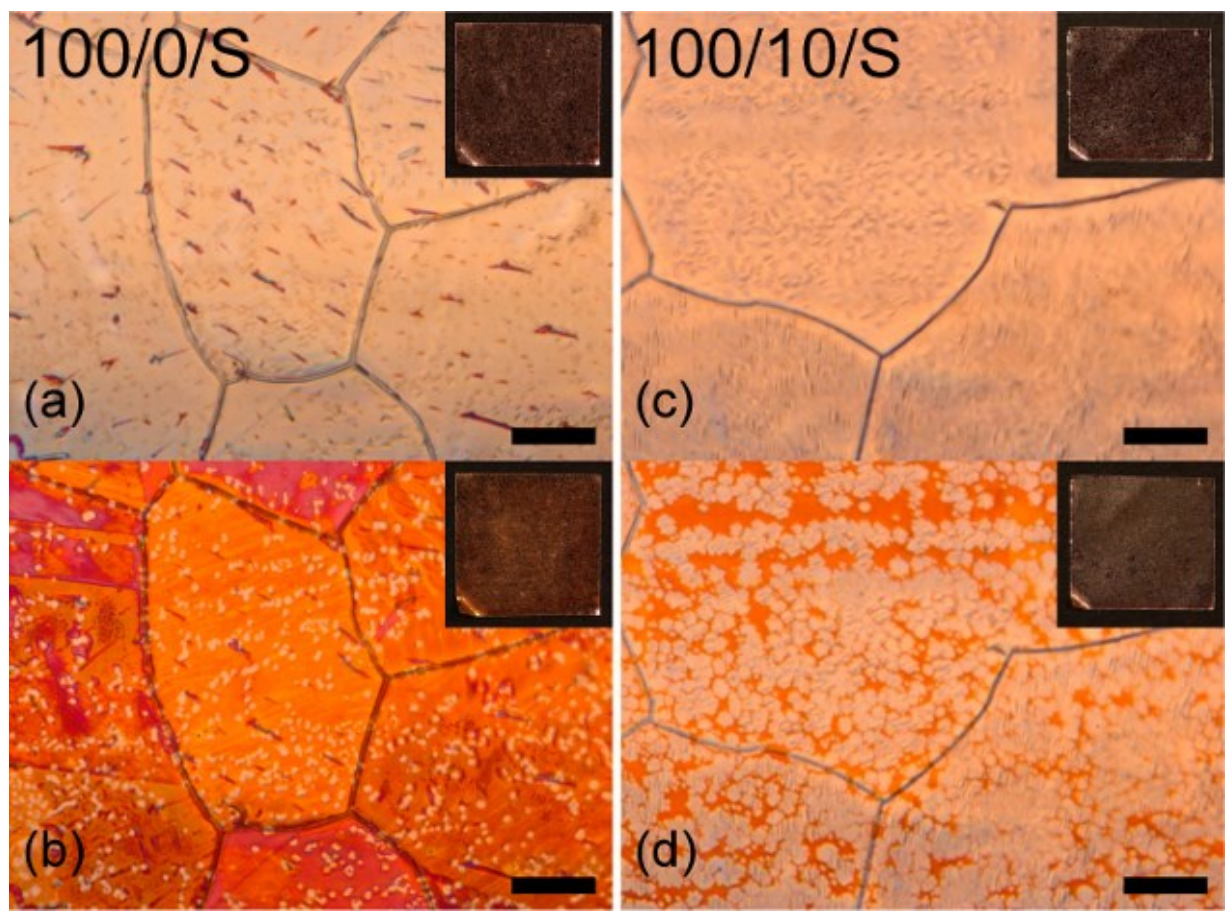

**Figure 1:** Optical microscopy pictures of samples 100/0/S and 100/10/S: (a) and (c) just after graphene growth; (b) and (d) after baking on a heating plate. The insets represent the photography of the corresponding whole copper foil (~1 cm$^2$). The scale bars are 20 μm.

To decrypt the first observations, the surface of three samples of type 100/0/S and three of type 100/10/S produced in different batches are analyzed by XPS. Some of them are examined in three different spots (spot size: 200 μm) and show excellent reproducibility across the sample. Core level spectra are recorded from carbon (C 1$s$), oxygen (O 1$s$), and copper (Cu 2*p*). In Fig. 2a, a typical survey scan shows that our samples are not contaminated, containing only, as one could expect, oxygen, carbon, and copper. Table 2 summarizes the corresponding atomic C and O concentrations, with their average (µ), standard deviation (σ), and coefficient of variation (σ/µ). In agreement with the OM pictures, the C concentration is drastically reduced by more than a factor of 2 on average for samples 100/0/S, while the O concentration increases nearly fourfold, evidencing a considerable oxidation. Moreover, for the 100/10/S samples, the small coefficient of variation for the C concentration (4%) reveals a weak variability in the C coverage. On the other side, the wide spreads on the degree of oxidation and on the C amount for the 100/0/S samples both betray a poor control of the atmosphere inside the reactor in the absence of $H_2$ during the cooling.

| | Atomic concentration [%] | | | | | | | | | | | |
|---|---|---|---|---|---|---|---|---|---|---|---|---|
| | 100/10/S | | | | | | 100/0/S | | | | | |
| | 1 | 2 | 3 | µ | σ | σ/µ | 1 | 2 | 3 | µ | σ | σ/µ |
| C | 69.4 | 65.5 | 72 | 69 | 2.7 | 0.04 | 30.1 | 38.4 | 28.4 | 32.3 | 4.4 | 0.14 |
| O | 7 | 12.4 | 7.4 | 9 | 2.5 | 0.27 | 26.1 | 37.7 | 37.6 | 33.8 | 5.4 | 0.16 |

**Table 2:** Carbon and oxygen atomic concentrations for three 100/10/S samples and for three 100/0/S samples. µ and σ correspond to the concentration mean value and standard deviation among the three samples, respectively.

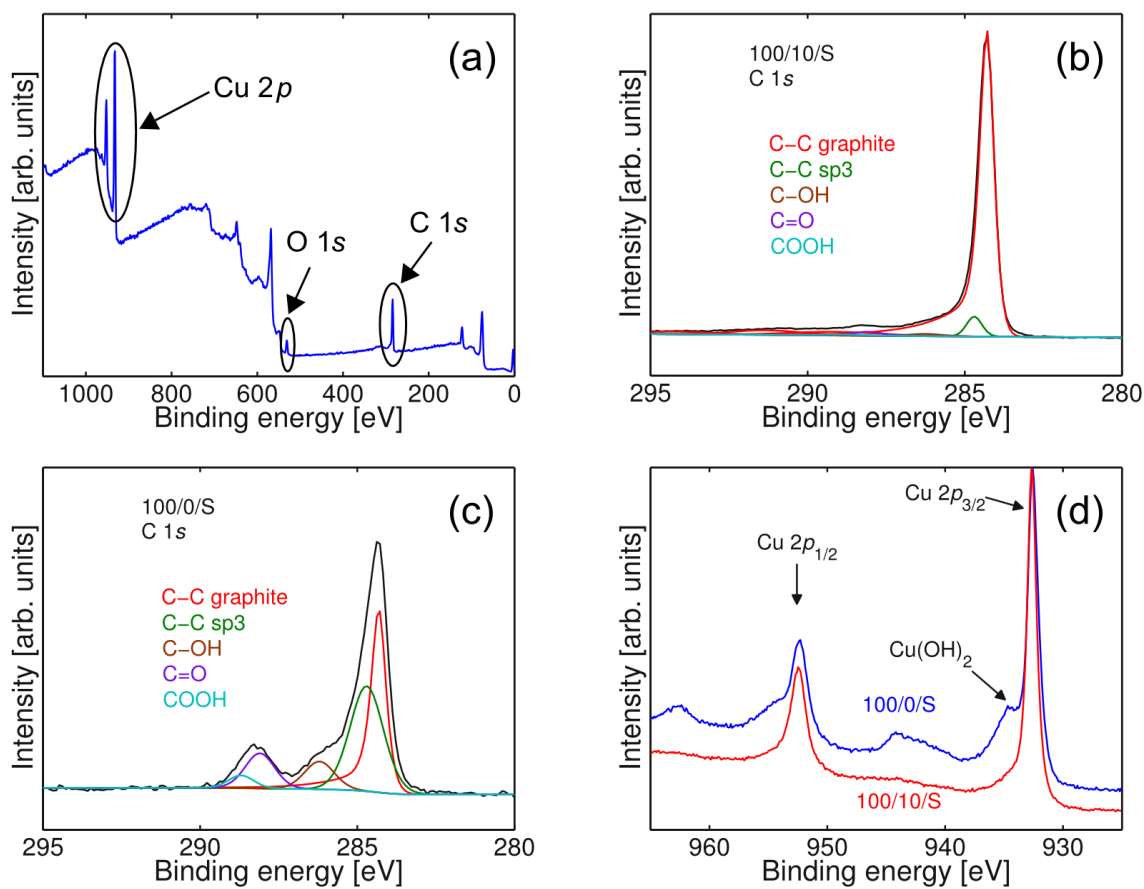

**Figure 2:** (a) Typical x-ray photoelectron spectroscopy scan survey showing carbon, copper and oxygen. (b) C 1s spectrum of sample 100/10/S and its fit. (c) C 1s spectrum of sample 100/0/S and its fit. (d) Cu 2p spectrum of samples 100/10/S and 100/0/S.

Fig. 2b and c display typical C 1s peaks of one of the 100/10/S and 100/0/S samples compared with a spectrum extracted from graphite. The C 1s peak of sample 100/10/S exhibits an asymmetric shape typical of graphitic sp2 carbon, with a binding energy (BE) of 284.3 eV and a full width at half maximum (FWHM) of 0.59 eV. It compares very well to the spectrum obtained from graphite, except for the FWHM which is slightly narrower for graphite (0.54 eV). The remainder of the spectrum can be adequately fit by four different contributions: C-C sp3 (284.7 eV), hydroxyl C-OH (286.2 eV), carbonyl C=O (288.1 eV), and carboxyl COOH (288.7 eV) groups (see Fig. 2b) [36]. The relative concentrations of the five types of carbon bonds are recapitulated in Table 3. For the 100/10/S samples, more than 90% of the C can be attributed to C-C sp2 (with a small dispersion of 1.8%, testimony of a good reproducibility), while the remainder (amorphous and oxidized carbon) could probably be associated to graphene flake edges (known to be chemically very active [32]), some contamination, etc. By contrast, the C-C sp2 fraction is smaller than 50% for the 100/0/S samples, which turn out to be heavily oxidized and amorphized. Once again, the large dispersions in the concentrations of amorphous and oxidized carbon are indicative of a poor control when $H_2$ is absent during the cooling.

|  | Carbon bonds relative concentration [%] | | | | | |
|---|---|---|---|---|---|---|
|  | 100/10/S | | | 100/0/S | | |
|  | 1 | 2 | 3 | 1 | 2 | 3 |
| C-C sp2 | 89.4 | 88 | 92.8 | 49 | 47.3 | 38.1 |
| C-C sp3 | 7.6 | 5.5 | 4.2 | 21.4 | 31.8 | 38.1 |
| C-OH | 1 | 2.9 | 1 | 14.8 | 8.5 | 9 |
| C=O | 0.9 | 1.9 | 1.6 | 4.2 | 8.4 | 11.3 |
| COOH | 1.1 | 1.7 | 0.4 | 10.6 | 4 | 3.5 |

**Table 3:** Carbon bonds relative concentrations for three 100/10/S samples and for three 100/0/S samples.

Alternatively, the Cu 2$p$ spectrum of both samples is shown in Fig. 2d. The following analysis relies on information extracted from references [29,37]. The Cu 2$p$ spectrum of sample 100/10/S is characteristic of metallic copper, with a Cu 2$p_{3/2}$ BE of 932.7 eV and a Cu 2$p_{1/2}$ peak about 20 eV higher in energy. The small oxygen atomic concentration should plausibly be distributed between "C-O" and "Cu-O" bonds (exposure of the copper surface to air after growth), even though the Cu 2$p$ does not indicate signs of it. The occurrence of CuO seems excluded (no high intensity shake-up peaks). By contrast, the Cu 2$p$ spectrum of sample 100/0/S is strongly affected. The signature of CuO can be clearly identified by the shake-up satellite peaks (two overlapping peaks at ~942.4 and 944.6 eV, and one at 963 eV). The shoulder (BE ~ 934.7 eV) on the Cu 2$p_{3/2}$ peak centered at 932.6 eV is relevant to Cu(OH)$_2$ (see the supplementary material for a discussion about its formation). The presence of Cu$_2$O is likely but is difficult to ascertain based solely on XPS analysis since its spectrum is very similar to metallic copper. A precise identification of the Cu compounds from the O 1$s$ peak and their proportions is difficult because of the overlap of C-O and Cu-O peaks but it is not essential to the present discussion.

The clear conclusion of the previous XPS investigation is that samples 100/0/S are considerably oxidized compared with samples 100/10/S. It is thus mandatory to flow H$_2$ during the slow cooling, which appears to have a protective effect on graphene against etching by oxidizing impurities at high temperatures. A recent paper indeed mentions nonuniform etching of graphene after exposure to air at temperatures greater than 400 °C [38]. Likewise, the growth of graphene under CH$_4$ alone turns out to be drastically inhibited by the concomitant oxidative action of O$_2$, the corresponding XPS results closely resembling those of 100/0/S samples. Another possibility to avoid the destruction of graphene during the cooling is to limit its exposure to oxidizing contaminants at high temperatures after growth by fast cooling (solution chosen by most authors, in conjunction with H$_2$ flow). The two copper foil pieces cooled rapidly under Ar flow with or without $H_2$ are also analyzed by XPS and both indeed show C 1$s$ spectra barely affected by oxidation or amorphization, looking very much alike the ones of the 100/10/S samples.

It is commonly admitted in the literature that $H_2$ has a twofold purpose [12]: acting both as a co-catalyst, promoting hydrocarbon decomposition and thereby graphene formation, and as an etchant reagent. In addition, $H_2$ has the important role to reduce the oxidized copper surface to enable the growth of graphene on top of it. In view of the experiments and deductions made in the recently published reference [26], it seems that $H_2$ does in fact not etch graphene, but that the etching is rather related to trace amounts of oxidizing impurities contained in the $H_2$ source. In our conditions, it appears that, even if oxidizing impurities stemming from both residual air and trace amounts in $H_2$ and $CH_4$, the overall effect of $H_2$ during the growth and during the cooling is very beneficial. In the absence of $H_2$, the deleterious effect of the oxidizing impurities manifests itself. It is very plausible that $O_2$ is the main oxidizing impurity. Forming gas, a few percent of $H_2$ mixed to Ar or $N_2$, is routinely used in metallurgy and integrated circuit technology to convert atmospheric $O_2$ into water in medium to high temperature processes, which is then evacuated with the gas flow. Thus, it is very likely that the role of $H_2$ during the growth and the cooling is the same in our case: to remove the oxidizing species from the furnace via reduction, thereby inhibiting their etchant effect. This does not mean that the etching by the oxidizing impurities is completely prevented in the presence of $H_2$ but that, at least, it is strongly impeded. We have conducted a supplementary experiment where the supply of $CH_4$ is suppressed for 30 min just after the growth stage and the temperature remains at 1000 °C (similar to Choubak *et al.* [26]). After that post-growth annealing, the copper foil is cooled rapidly under $H_2$ as usual. Expectedly, the resulting graphene sheet proves out heavily etched after the annealing (see Fig. S3), betraying the occurrence of residual oxidizing contaminants in the furnace atmosphere etching the graphene film in spite of the presence of $H_2$. In the same line, graphene growth is drastically restrained without $H_2$ because, in that case, oxidizing impurities are not removed from the reactor and graphene etching is predominant over graphene growth, resulting in discontinuous and heavily oxidized, etched graphene films, as illustrated by Fig. S1b. Very similar conclusions are drawn in [26] for graphene grown solely with $CH_4$, because of insufficient gas purity. In the present approach, we tolerate the presence of oxidizing impurities (and in fact, they cannot be completely avoided in our vacuum-free system) but their harmful effect is substantially impaired via reduction with $H_2$, while Choubak *et al.* [26] must remove the contaminants as much as possible to avoid any etching.

It is thus evidenced here that $H_2$ has a protective role on graphene, by counteracting the presence of oxidizing impurities. The question now is: where do these contaminants come from? The system that we use is a very simple one: a quartz tube connected on one side to gas cylinders through mass flows and on the other side to the gas exhaust, without any pumping system. The quartz tube can be inserted and extracted easily from the hot-wall furnace. An obvious potential source of $O_2$ is air leakage in the gas circuit. After a thorough investigation of leakages in our system by means of a mass-spectrometer leak detector, that hypothesis can be excluded. Alternatively, the quartz tube is kept under ambient air most of the time. When the sample is introduced on the quartz boat and the tube is sealed, it is full of air. In our growth process, a high Ar debit of 2000 sccm is first flown for 15 min at room temperature to flush air inside the quartz tube. Next, the tube is inserted into the furnace at 700 °C under an additional $H_2$ flow (100 sccm) and the copper foil is annealed at 1000 °C for 1 h. By doing so, air is only diluted in Ar, meaning that there is always some residual $O_2$ mixed with the other gases inside the reactor. A way to improve the process would be to add a pumping system to perform several cycles of pumping and purges [10], at the expense of simplicity and cost. Another source of contamination which cannot be ignored may originate from trace amounts of oxidizing impurities in the gas cylinders themselves. In effect, the $H_2$ and, more particularly, the $CH_4$ bottles are of low purity (99.9 and 99.5% purity, respectively). Still, it is quite difficult to discriminate between these two possible sources of contaminants. One possibility to make a distinction could be to make use of gas purifiers for both $H_2$ and $CH_4$. On the other hand, water is also known to be a carbon etchant at high temperatures [39,40]. The presence of water in the reactor cannot be excluded. Concerning moisture contained in the air, we believe that heating at high temperature should readily remove water at the early stage of the synthesis process, under a relatively important Ar flow. However, trace amounts of water are most certainly contained in the $H_2$ and $CH_4$ canisters. Finally, a growth following a 3-h-long Ar initial purge at 2000 sccm at room temperature (instead of the usual 15-min-long one) is also tested but likewise proved to be unsuccessful unless $H_2$ is introduced during natural cooling (see Fig. S4). Consequently, in such a rudimentary system, the occurrence of oxidizing species seems inevitable.

Fig. 3a illustrates a SEM top view of graphene domains on copper (sample 100/10/F), revealing their hexagonal shape. Fig. 3b gives a close-up view of a typical hexagon (~3 μm from vertex to vertex). The ripples observed previously by OM can be clearly seen on several hexagons. Some domains seem monolayer and others few-layer, with the nucleation seed visible in the center, with random orientations. Raman spectroscopy is used to check for the structural quality of the graphene hexagonal domains and the number of layers [41]. The possible degradation of graphene quality due to residual $O_2$ without flowing $H_2$ during the fast cooling is also considered. Three sorts of samples are investigated: 100/10/S, 100/10/F, and 100/0/F. Two samples of type 100/10/F from two different batches are analyzed to assess the reproducibility of the growth and transfer processes. The inset to Fig. 4 displays an OM image of a 100/10/F sample after transfer on 300-nm-thick $SiO_2$. The Raman measurements are limited to the hexagons with the weakest contrast (see the highlighted hexagon in the inset to

Fig. 4). For each sample, 10 hexagons (at least 2-µm-wide to avoid edges) scattered all over the SiO$_2$ piece are inspected. Fig. 4 shows three typical Raman spectra, one for each type of sample. Characteristic features of graphene, the so-called G band (~1590 cm$^{-1}$), G* band (~2450 cm$^{-1}$) and G' band (~2690 cm$^{-1}$), are visible in the Raman spectra [41,42]. As it can be seen, the disorder-related D band (~1350 cm$^{-1}$) is not observable, sign of excellent structural quality. The non-perturbed G band is usually around 1582 cm$^{-1}$. The slight upshift is most probably due to residual strain from the copper substrate and unintentional doping [43,44,45] (possibly coming from traps in the SiO$_2$ substrate, from insufficient rinsing after copper etching, from residues of PMMA after removal in acetone, moisture in the air, and so on). Consequently, within the limits of this study, it turns out that the cooling rate has no peculiar influence on the quality of graphene (at least with H$_2$) and that residual O$_2$ does not degrade graphene during the cooling if the cooling is performed fast enough. This is most likely the reason why most authors make use of fast cooling. Moreover, the transfer procedure does neither introduce any additional defects. The average over the four samples of the G band position, G' band position, FWHM of the G' band, and *I*(G')/*I*(G) ratios amount to 1589±3 cm$^{-1}$, 2687±3 cm$^{-1}$, 35±2 cm$^{-1}$, and 1.6±0.2, respectively (see details in Table 4). All of these results confirm that the corresponding hexagons are monolayer graphene [10,42] and testify of the very good reproducibility of both the APCVD growth and of the transfer process.

|  | Raman parameters | | | | | | | |
|---|---|---|---|---|---|---|---|---|
|  | 100/10/S | | 100/10/F #1 | | 100/10/F #2 | | 100/0/F | |
|  | μ | σ | μ | σ | μ | σ | μ | σ |
| $I(G')/I(G)$ | 1.5 | 0.1 | 1.7 | 0.1 | 1.5 | 0.2 | 1.9 | 0.1 |
| G [cm$^{-1}$] | 1590 | 2 | 1590 | 2 | 1589 | 3 | 1590 | 1 |
| G' [cm$^{-1}$] | 2690 | 2 | 2686 | 2 | 2687 | 4 | 2690 | 2 |
| FWHM G' [cm$^{-1}$] | 36 | 1 | 35 | 1 | 36 | 2 | 36 | 1 |

**Table 4:** Summary of the parameters extracted from Raman spectra for samples 100/10/S, 100/10/F, and 100/0/F. μ and σ correspond to the Raman parameter mean value and standard deviation among the four samples, respectively.

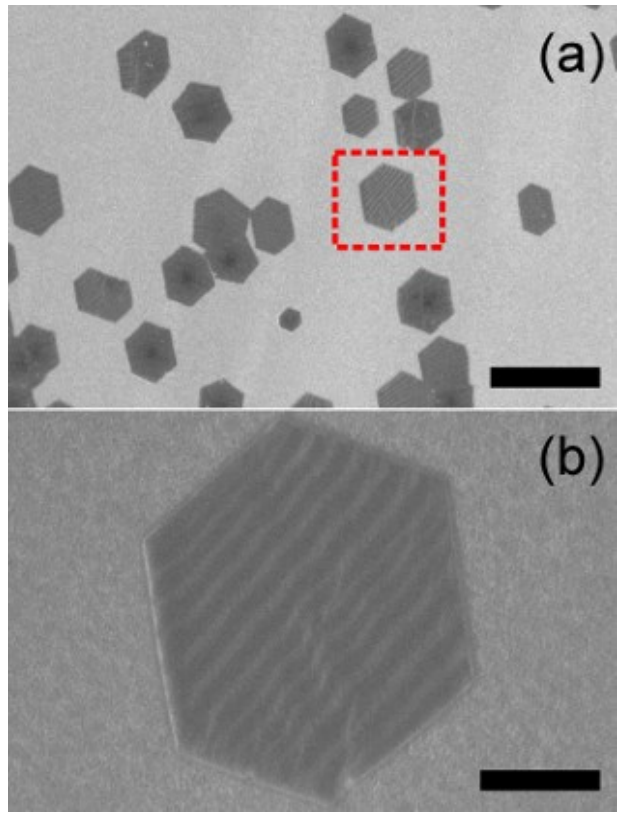

**Figure 3:** (a) Scanning electron microscopy image of graphene hexagonal flakes on copper foil. (b) Zoom on a monolayer graphene hexagon. The scale bars are 5 and 1 μm, respectively.

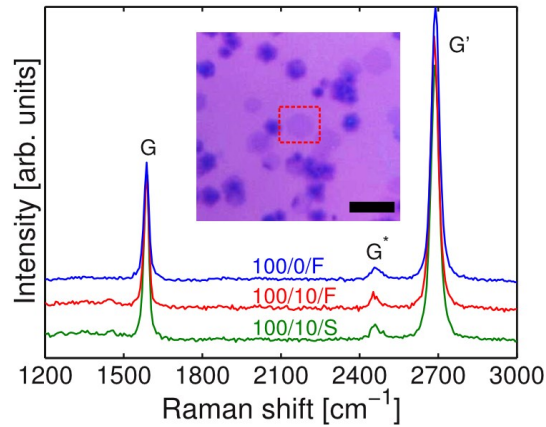

**Figure 4:** Raman spectra of the 100/10/S, 100/10/F, and 100/0/F samples. Inset: optical microscopy image of graphene hexagons transferred onto $SiO_2$. A monolayer flake is emphasized by the red dashed rectangle. The scale bar is 5 μm.

## 4. Conclusion

The growth of graphene on copper foils with $CH_4$ by APCVD in a vacuum-free furnace is explored. More precisely, we concentrate on the consequences of $H_2$ suppression during the growth or the cooling. First, OM and XPS analyses evidence that slow cooling of the copper foil without $H_2$ results in low carbon coverage with heavily oxidized and amorphized graphene. By contrast, the addition of a small amount of $H_2$ or fast cooling leads to appreciable graphene coverage. A likely explanation for this observation is the seemingly inevitable presence of residual oxidizing impurities from ambient air in the growth atmosphere and from insufficiently pure gases, which strongly damages graphene upon too long exposure at high temperatures. Likewise, graphene formation is drastically inhibited if $H_2$ is not admitted during the growth step. In consequence, $H_2$ must be present all along the process to prevent a re-oxidation of the copper surface during growth and also to protect graphene from etching during slow cooling. Fast cooling with $H_2$ is the safest way to cool down the copper foil. In the best conditions, micrometer-sized graphene hexagons are formed. Raman spectroscopy and SEM confirm that these domains are monolayer or few-layer. The monolayer hexagons are found to be of excellent structural quality, as testified by the absence of D band in the Raman spectra. In conclusion, even though the presence of residual oxidizing contaminants in the growth atmosphere is inevitable in vacuum-free reactors, it is not an obstacle to the formation of virtually defect-free graphene under the appropriate conditions.

## Acknowledgements

This work is supported by the Belgian Fund for Scientific Research (FRS-FNRS) under FRFC contract "Chemographene" (convention N°2.4577.11). B. Hackens and J.-F. Colomer are supported by the Belgian Fund for Scientific Research (FRS-FNRS) as Research Associates. A. Felten is supported by the Belgian Fund for Scientific Research (FRS-FNRS) as postdoctoral researcher. C. N. Santos acknowledges financial support from the ARC "StressTronics" project. The authors acknowledge N. Moreau for useful discussions.

# Supplementary Material

**The influence of oxidizing impurities on the synthesis of graphene by atmospheric pressure chemical vapor deposition**

Nicolas Reckinger[a], Alexandre Felten[a], Cristiane N. Santos[b], Benoît Hackens[b], and Jean-François Colomer[a]

[a]Research Center in Physics of Matter and Radiation (PMR), University of Namur, Rue de Bruxelles 61, B-5000 Namur, Belgium

[b]Université catholique de Louvain (UCL), Institute of Condensed Matter and Nanosciences (IMCN), Nanophysics Division (NAPS), Chemin du Cyclotron 2, 1348 Louvain-la-Neuve, Belgium

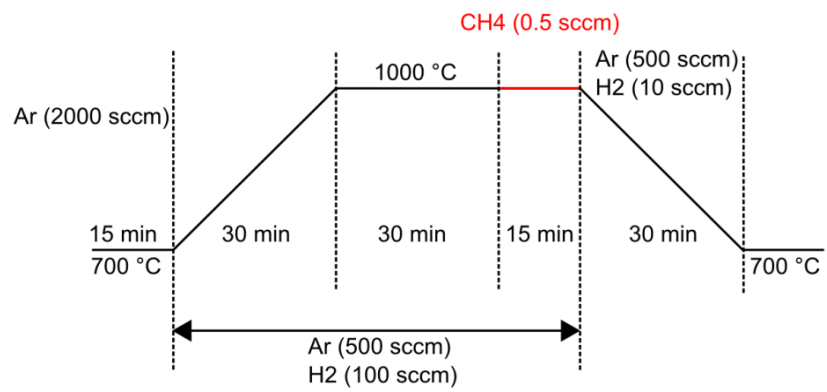

**Figure S1:** Furnace temperature versus time graph for the synthesis process, with the corresponding Ar, $H_2$, and $CH_4$ flows.

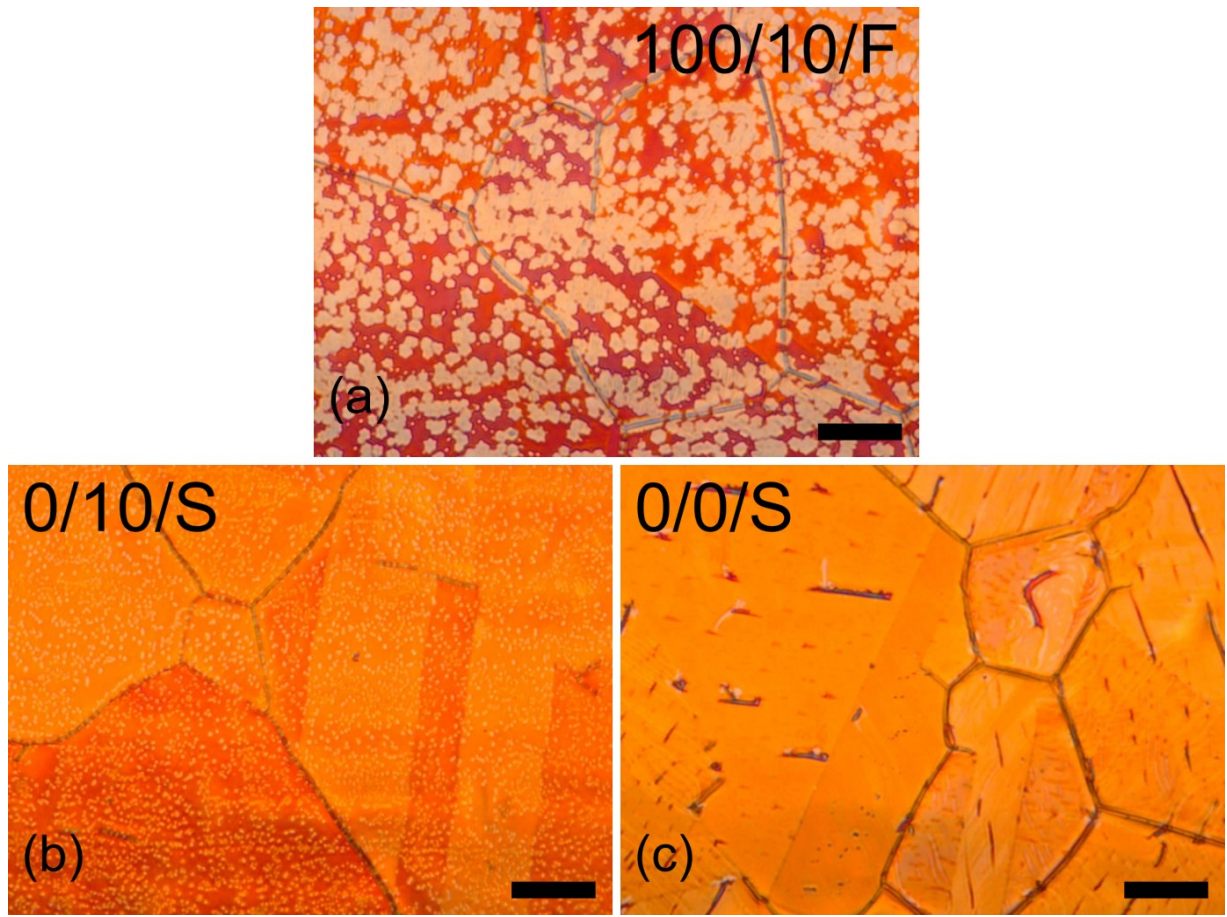

**Figure S2:** Optical microscopy pictures of samples 100/0/F (a), 0/10/S (b), and 0/0/S (c) after baking on a heating plate. The scale bars are 20 μm.

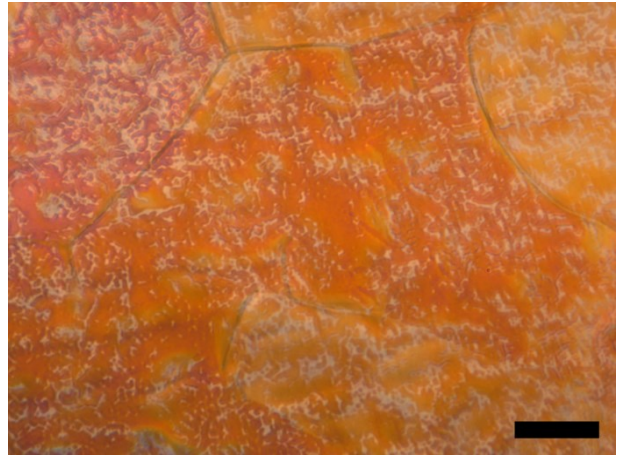

**Figure S3:** Optical microscopy picture of a sample of type 100/10/F with a 30-min-long post-growth annealing, after baking on a heating plate. The scale bar is 20 µm.

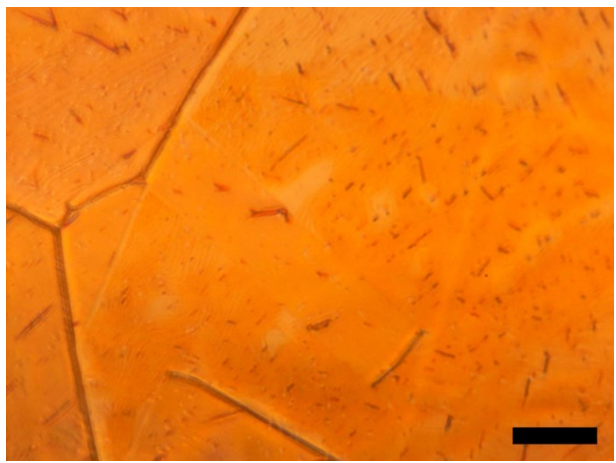

**Figure S4:** Optical microscopy picture of a sample of type 100/0/S with a 3-h-long pre-growth purge, after baking on a heating plate. The scale bar is 20 µm.

Regarding the formation of copper hydroxide, the values of the redox potential of the $Cu^{2+}/Cu$ (0.342) and $H_2O/H_2$ (-0.828) show that the reaction of water with copper to form copper hydroxide is not possible due to the smaller redox potential value of water. This means that water cannot oxidize metallic Cu. This fact is well known in metallurgy, where atmospheric oxygen is converted into water during annealing at high temperature by addition of hydrogen to avoid the formation of copper oxide.

$Cu^{2+} + 2e^- \leftrightarrow Cu$: redox potential = 0.342

$H_2 + 2OH^- \leftrightarrow 2H_2O + 2e^-$: redox potential = -0.828

Copper hydroxide comes from the first steps of copper oxide reduction by hydrogen as explained by reference [1], studying the reduction of copper oxide thin films with a hydrogen plasma.